\documentclass[12pt]{article}
\usepackage{amssymb,amsmath,epsfig}
\usepackage{graphicx}
\usepackage{xcolor}
\allowdisplaybreaks
\begin{document}
\title{\textbf{A Non-Singular Cosmic Bounce in $\mathcal{F}(Q)$ Gravity:
A Reconstruction and Phase Space Analysis}}
\author{Nusrat Fatima\thanks{nusratfatimaliaqat@gmail.com} ~and
M. Sharif\thanks{msharif.math@pu.edu.pk}\\
Department of Mathematics and Statistics, The University of Lahore,\\
1-KM Defence Road Lahore-54000, Pakistan.}

\date{}
\maketitle

\begin{abstract}
The key focus of this research work is the analysis of a
non-singular cosmic bounce in the context of $\mathcal{F}(Q)$
gravity, with $Q$ representing the non-metricity scalar. We consider
the gravitational Lagrangian $\mathcal{F}(Q)=Q+\psi Q^n$ in the
presence of a modified Chaplygin-type matter source with a flat
Friedmann-Robertson-Walker spacetime and a perfect matter
distribution. In order to proceed, we adopt two approaches: a
reconstruction approach using scale-factor ansatz and analysis of a
two-dimensional autonomous dynamical system. Our results imply that
the geometry coupling parameter $\psi$ plays an important role and
causes geometrical repulsion for violating the null energy
condition. A numerical scan of the $(\rho_0,\psi)$ parameter space
suggests that there is a critical energy density for the bounce to
take place, below which the universe evolves towards the standard
singular cosmological solution. Moreover, the effective equation of
state slowly evolves towards the de Sitter value
($\omega_{eff}\rightarrow -1$) and the squared sound speed is
bounded by the stability and causality condition ($0\leq
C_s^2\leq1$). It is found that $\mathcal{F}(Q)$ gravity provides a
coherent geometric picture of the non-singular bouncing universe in
accordance with the cosmic accelerated expansion.
\end{abstract}

\section{Introduction}

Modern understanding of cosmic evolution is primarily based on
General Relativity (\textit{GR}) together with the highly successful
$\Lambda$CDM cosmological background \cite{1}-\cite{4}. The
scientific community describes a universe that emerged from an
extremely hot and dense primordial state and has undergone
continuous expansion over cosmic time. Although this framework
successfully explains a wide range of cosmological observations, it
does not give a complete explanation of the earliest cosmic stages.
One of its most fundamental shortcomings is the prediction of an
initial spacetime singularity. According to the singularity theorems
established by Penrose and Hawking \cite{1}, any expanding universe
governed by \textit{GR} and filled with ordinary matter inevitably
evolves backward to a state of infinite density and divergent
spacetime curvature. At this singular point, the classical
description of gravity becomes invalid, indicating that \textit{GR}
is unlikely to remain applicable in the extreme high energy
conditions (\textit{ECs}) of the primordial universe.

The presence of this singularity has motivated extensive efforts
towards constructing alternative cosmic scenarios capable of
providing a regular cosmic origin. Among the proposed possibilities,
non-singular bouncing cosmology has attracted considerable
attention. Instead of emerging from an initial singular state, the
universe is assumed to undergo a contracting epoch followed by a
smooth transition into the present expanding phase. Such cosmic
bounce refers to a transition from contraction to expansion over the
time evolution of the Hubble parameter, characterized by the
conditions $H=0$ and $\dot{H}>0$. In \textit{GR} \cite{6}-\cite{8},
the Raychaudhuri equation prohibits this change from contraction to
expansion when the matter sector obeys the null energy condition
(\textit{NEC}), $\rho+p\geq0$. Consequently, realizing a stable
cosmic bounce generally requires either the presence of
unconventional matter components capable of violating the
\textit{NEC} or modifications to the underlying gravitational
theory.

Modified theories of gravity provide a compelling geometric
mechanism for generating an effective violation of the \textit{NEC}
while avoiding the pathological ghost instabilities often associated
with exotic matter fields. In this direction, Mironov \textit{et al}
\cite{10} presented a comprehensive review of non-singular
early-universe scenarios, including cosmological bounce and genesis
models formulated within Horndeski scalar-tensor gravity and its
extensions. Their work summarizes the theoretical progress achieved
in constructing stable alternatives to the standard Big Bang
picture. Furthermore, numerous investigations of bouncing cosmos
have been carried out in different modified gravity frameworks and
matter models \cite{11}-\cite{18}.

There is a broad class of modified gravitational theories, including
$\mathcal{F}(\mathcal{R})$ gravity ($\mathcal{R}$ is the curvature
of spacetime) \cite{19}-\cite{21}, $\mathcal{F}(T)$ gravity (where
$T$ represents the torsion) \cite{22}-\cite{25},
$\mathcal{F}(\mathcal{R},\mathbb{T})$ gravity ($\mathbb{T}$
expresses the trace of stress energy tensor) \cite{25a} and other
geometric extensions of \textit{GR}. Recently, $\mathcal{F}(Q)$
theory has received considerable attention due to its rich
cosmological phenomenology and the additional coupling between
matter and spacetime geometry, making it a versatile framework for
investigating the dynamics of the early universe. This framework was
first introduced by Jimenez \textit{et al} \cite{26}, where gravity
is entirely defined by non-metricity $Q$ while both curvature and
torsion are identically zero. By generalizing the gravitational
function from $Q$ to an arbitrary function $\mathcal{F}(Q)$, the
theory introduces additional gravitational contributions that can
influence both the cosmic accelerated expansion and its early
evolution without requiring exotic matter components. Furthermore,
in contrast to several other modified gravity theories, the field
equations of this theory retain their second-order structure,
rendering the formalism comparatively tractable and avoiding the
higher-order derivatives that frequently arise in curvature-based
gravitational extensions.

The study of bouncing cosmology in the context $\mathcal{F}(Q)$
gravity has attracted considerable importance from researchers due
to its distinctive and intriguing properties. Lazkoz \textit{et al}
\cite{2a} studied the cosmological and observational constraints of
this gravity and demonstrated that the accelerated expansion can
emerge intrinsically from the underlying cosmic geometric structure.
Mandal \textit{et al} \cite{2d} examined the \textit{ECs} and
constrained the model variables using the present values of the
cosmological parameters to assess the viability of their
cosmological models within this theory. They also carried out an
extensive cosmographic analysis within the same theoretical
framework \cite{2c}. Bajardi \textit{et al} \cite{2e} investigated
the Hamiltonian and ADM formalisms to determine the cosmic wave
function in this framework. Mandal \textit{et al} \cite{2f} examined
bouncing scenarios using two distinct Lagrangian forms of
$\mathcal{F}(Q)$ through a perturbative approach and found that the
perturbation term attains a significantly large value near the
bounce and subsequently approaches zero. For a comprehensive
analysis of the geometric foundations of this gravity in the context
of Friedmann-Robertson-Walker (FRW) cosmology, we refer to the work
of Sharif \textit{et al} \cite{27}-\cite{30}. Motivated by this
theoretical foundation, recent investigations have concentrated on
particular functional forms capable of accounting early-time
inflationary dynamics and late-time cosmic acceleration
\cite{31}-\cite{33}.

In this study, we investigate a polynomial extension of symmetric
teleparallel gravity described by the functional form
$\mathcal{F}(Q)=Q+\psi Q^{n}$, where $\psi$ and $n$ are arbitrary
constant model variables that characterize the departure from
\textit{GR}. The non-metricity correction $Q^{n}$ becomes important
under the high \textit{ECs} of the early universe. It modifies the
gravitational dynamics and therefore offers a useful ground for
studying non-singular cosmic bounce. For the cosmic matter content,
we employ the Chaplygin gas equation of state (\textit{EoS}), which
can effectively represent both dark matter and dark energy within a
unified framework while obeying the conventional condition
$\rho+p>0$. Consequently, the violation of \textit{NEC}, which is
essential for realizing a cosmological bounce, originates entirely
from the geometric modifications introduced through the
non-metricity sector rather than from the matter content itself.
Owing to this interpolating behavior, the Chaplygin gas offers a
unified description of the dominant cosmological components and
serves as a suitable matter candidate for investigating the complete
history of cosmic evolution. Furthermore, unlike phantom or
scalar-field models, the Chaplygin gas does not introduce additional
dynamical degrees of freedom or ghost instabilities, making it an
attractive matter source for studying viable bouncing solutions in
modified gravity \cite{34}-\cite{37}.

The structure of this article is organized as follows. Section
\textbf{2} presents the fundamental formulation of the
$\mathcal{F}(Q)$ gravitational framework and derives the
corresponding modified Friedmann equations. The conditions required
for achieving a non-singular cosmological bounce are also
established within this geometric framework. In Section \textbf{3},
the cosmological dynamics are investigated through two complementary
approaches. The first employs a symmetric bouncing scale factor to
reconstruct the cosmological evolution, whereas the second examines
the underlying dynamics using an autonomous dynamical system. The
physical acceptability of both scenarios is subsequently assessed
through the analysis of the \textit{ECs}, the stability criterion
based on the squared sound speed and the corresponding phase-space
trajectories. A detailed comparison of the two approaches, together
with a discussion of their cosmological implications and consistency
with non-singular bouncing models, is presented in Section
\textbf{4}. Finally, the principal results of the study and their
significance for early-universe cosmology are summarized in Section
\textbf{5}.

\section{Theoretical Foundation of $\mathcal{F}(Q)$ Gravity}

Following the discussion of the physical motivations, we now present
the theoretical framework employed in this work. We start with the
gravitational action of $\mathcal{F}(Q)$ gravity generalized from
the symmetric teleparallel gravity by swapping the non-metricity
scalar $Q$ with an arbitrary function $\mathcal{F}(Q)$. This action
serves as the basis for deriving the modified field equations used
throughout our analysis and is defined as~\cite{26}
\begin{equation}\label{1}
S=\int\left[\frac{1}{2\kappa}{\mathcal{F}(Q)}
+\mathcal{L}_{m}\right]\sqrt{-g}\,d^{4}x.
\end{equation}
Here $\kappa=1$ represents the coupling constant, $g$ denotes the
determinant of the metric tensor and $\mathcal{L}_{m}$ represents
the matter lagrangian density. By varying the action with respect to
the metric tensor $g_{\mu\nu}$, the associated field equations are
obtained as
\begin{equation}\label{2}
-\frac{2}{\sqrt{-g}} \nabla_{\eta} \left(\sqrt{-g} {\mathcal{F}_{Q}}
P^{\eta}_{\mu\nu}\right) -\frac{1}{2}g_{\mu\nu}\mathcal{F}(Q)
-\mathcal{F}_Q \left(P_{\mu\eta\beta} Q_{\nu}^{\ \ \eta\beta} -
2Q_{\mu}^{\ \ \eta\beta} P_{\beta\eta\nu} \right) =
\mathcal{T}_{\mu\nu},
\end{equation}
where $\mathcal{F}_{Q} \equiv \partial \mathcal{F}/\partial Q$
denotes the derivative of the function $\mathcal{F}(Q)$ with respect
to the non-metricity scalar  $Q$. In terms of superpotential tensor,
the non-metricity scalar is given by
\begin{equation}\label{3}
Q=-Q_{\eta\mu\nu}P^{\eta\mu\nu}
=-\frac{1}{4}\big[-{Q}^{\eta\mu\nu}{Q}_{\eta\mu\nu}
+2{Q}^{\eta\mu\nu}{Q}_{\nu\eta\mu}-2{Q}^{\eta}
\tilde{{Q}}_{\eta}+{Q}^{\eta}{Q}_{\eta}\big],
\end{equation}
where
\begin{equation}\label{4}
Q_{\eta\mu\nu}=\nabla_{\eta}g_{\mu\nu}\neq0,\qquad
Q_{\eta}=Q_{\eta~~\mu}^{~~\mu}, \qquad \tilde Q_{\eta}=Q^{\mu}_{\ \
\eta\mu}.
\end{equation}
Furthermore, the superpotential tensor can be given as
\begin{equation}\label{5}
P^{\eta}_{\ \mu\nu} = -\frac14Q^{\eta}_{\ \mu\nu} +\frac12Q_{(\mu\
\nu)}^{\ \ \eta} +\frac14 \left(Q^{\eta}-\tilde
Q^{\eta}\right)g_{\mu\nu} -\frac{1}{4} (\delta^{\eta}_{\
\mu}Q_{\nu}+\delta^{\eta}_{\ \nu}Q_{\mu}).
\end{equation}
The matter distribution is characterized by the energy-momentum
tensor. In the present work, we consider an isotropic matter
configuration, whose energy-momentum tensor is given by
\begin{equation}\label{6}
\mathcal{T}_{\mu\nu} =(\rho+p)u_{\mu}u_{\nu} +pg_{\mu\nu}.
\end{equation}
Here, $u_{\mu}$ denotes the four-velocity of the cosmic fluid, $p$
is the fluid pressure, $\rho$ is the energy density.

For the cosmological background, we adopt the FRW spacetime, which
takes the form
\begin{equation}\label{7}
ds^{2} = -dt^{2} +a^{2}(t) (dx^{2}+dy^{2}+dz^{2}),
\end{equation}
here, $a(t)$ is the scale factor. From Eqs.\eqref{3} and \eqref{7},
we obtain the non-metricity scalar as
\begin{equation}\label{8}
Q=6H^{2},
\end{equation}
$H$ is the Hubble parameter. The resulting field equations are
obtained as
\begin{eqnarray}\label{9}
3H^{2} &=& \frac{1}{2 \mathcal{F}_Q}\left(
\rho-\frac{\mathcal{F}(Q)}{2} \right),
\label{10}\\
2\dot{H}+3H^{2}&=&\frac{1}{2 \mathcal{F}_Q}\left( p+\frac{
\mathcal{F}(Q)}{2}\right)-2\mathcal{F}_{Q Q}H.
\end{eqnarray}
To investigate the cosmic bounce in $\mathcal{F}(Q)$ gravity, it is
necessary to specify the functional form of the gravitational
Lagrangian. Inspired by the analysis of Harko \textit{et
al}~\cite{38}, we adopt the polynomial model \cite{38a}
\begin{equation}\label{11}
\mathcal{F}(Q)=Q+\psi Q^{n},\quad n\neq0.
\end{equation}
This functional form has been extensively used in the literature to
investigate the cosmological behavior of modified symmetric gravity.
The modified Friedmann equations corresponding to the particular
functional form are obtained as
\begin{eqnarray}
3H^{2} &=& \frac{1}{2\left(1+n\psi Q^{\,n-1}\right)} \left(
\rho-\frac{Q+\psi Q^{n}}{2} \right) =\rho_{\rm eff},
\label{12}\\
2\dot{H}+3H^{2} &=&- \frac{1}{2\left(1+n\psi Q^{\,n-1}\right)}
\left( p+\frac{Q+\psi Q^{n}}{2} \right) \nonumber\\\nonumber\\ &&
-(n-1)n\psi Q^{\,n-2}H = -p_{\rm eff}. \label{13}
\end{eqnarray}
Combining the above Friedmann equations yields the modified
Raychaudhuri equation, which defines the evolution of the Hubble
parameter and plays a key role in the analysis of non-singular
cosmic bounce
\begin{equation}\label{14}
\dot{H}=-\frac{1}{2}(\rho_{\emph{eff}}+p_{\emph{eff}})=-\frac{1}{2}
\bigg[\frac{\rho+p}{2\left(1+n\psi Q^{\,n-1}\right)}+2n(n-1)\psi
Q^{\,n-2}\bigg].
\end{equation}

In the framework of FRW cosmology, a modified Chaplygin gas model
has been proposed \cite{39}. This specific model has an early stage
of radiation and is defined by an \textit{EoS}:
\begin{equation}\label{15}
p(t)=A\rho-\frac{B}{\rho^{\gamma}}, \qquad 0\leq\gamma\leq1,
\end{equation}
where $A$ and $B$ are positive constants. When $B=0$, the modified
Chaplygin gas reduces to the perfect fluid \textit{EoS}, $p=A\rho$.
For $A=0$, it reduces to the generalized Chaplygin gas \textit{EoS}.
The two quantities in the \textit{EoS} become comparable when the
pressure vanishes, i.e., $p=0$. In this scenario, the fluid has
pressureless density $\rho_0$, associated to some scale factor
$a_0$,
\begin{equation}\label{16}
\rho_0=\rho^{\,\gamma+1}(a_{0})=\frac{B}{A}.
\end{equation}
In order to meet the bounce condition $\dot{H}>0$ at $H=0$, the
expression on the right of Eq.\eqref{14} must be positive. This
gives
\begin{equation}\label{17}
\frac{\rho+p}{2\left(1+n\psi Q^{\,n-1}\right)}+2n(n-1)\psi
Q^{\,n-2}<0.
\end{equation}
In a universe with Chaplygin gas, the matter sector usually obeys
the inequality  $\rho+p>0$. Thus, a non-singular cosmic bounce is
only possible when the gravitational effects dominate the cosmic
evolution
\begin{equation}\label{18}
\frac{\rho\big(1+A\big)-\frac{B}{\rho^\gamma}}{2\left(1+n\psi
Q^{\,n-1}\right)}+2n(n-1)\psi Q^{\,n-2}<0.
\end{equation}
This derivation indicates that the coupling parameter $\psi$ has a
key role in the bouncing dynamics. In particular, when $\psi>0$, the
correction term $\psi Q^{n}$ becomes increasingly significant at
high energy densities, generating an effective repulsive
gravitational contribution. This geometric effect can lead to the
violation of the \textit{NEC}, thereby enabling a cosmic bounce.

A non-singular cosmological bounce occurs at a time $t=t_b$ when the
Hubble parameter disappears and transits from negative to positive
regime. Thus, the required conditions for a bounce are
\begin{equation}\label{19}
H(t_b)=0, \qquad \dot{H}(t_b)>0.
\end{equation}
Since the non-metricity scalar is related to the Hubble parameter
through $Q=6H^2$, the bounce point also satisfies $Q(t_b)=0$. From
the modified Raychaudhuri equation (\ref{14}) and for the choice
$n=2$, its value at the bounce becomes
\begin{equation}\label{20}
\dot{H}_b=-\frac{1}{4}(\rho_b+p_b).
\end{equation}
Using Eq.\eqref{15}, the above relation takes the form
\begin{equation}\label{21}
\dot{H}_b =-\frac{1}{4} \left[ (1+A)\rho_b-\frac{B}{\rho_b^\gamma}
\right].
\end{equation}
Consequently, the condition $\dot{H}_b>0$ requires
\begin{equation}\label{22}
\rho_b+p_b<0,
\end{equation}
Therefore, for the present model, the occurrence of the bounce is
determined by the effective dynamics of the Chaplygin gas matter
together with the non-linear $\mathcal{F}(Q)$ corrections, rather
than by imposing a particular sign of the coupling parameter $\psi$.

\section{Kinematic Reconstruction and Cosmic Evolution}

As the basic criteria for a non-singular bounce have been
established in the previous section, we now examine the bouncing
behavior in the context of the polynomial model,
$\mathcal{F}(Q)=Q+\psi Q^{n}$. First, we employ the reconstruction
technique by choosing an appropriate scale factor ansatz to study
the occurrence of a non-singular bounce. Next, we investigate the
cosmological evolution through an autonomous dynamical system to
evaluate the stability and dynamical properties of the resulting
bouncing solutions.

\subsection{Model I: Reconstruction via Scale Factor Ansatz}

For the present model, we adopt a scale factor that represents a
smooth and symmetric evolution from the contracting universe to an
expanding one. The chosen ansatz is \cite{39a}
\begin{equation}\label{23}
a(t)=a_{0}\left(1+t^{2}\right)^{\frac{h}{2}},
\end{equation}
where $a_{0}$ denotes the value of the scale factor at the bounce
epoch ($t=0$), while $h$ is an arbitrary constant that controls the
expansion rate. The scale factor given in Eq.(\ref{23}) is selected
because of its desirable physical and mathematical features. It
remains finite and positive throughout the cosmic evolution, with
$a(0)=a_{0}\neq0$, thereby avoiding the initial singularity. This
chosen formulation of the scale factor guarantees a smooth and
continuous shift during the bounce, preserving a finite and non-zero
value at the bouncing epoch. This formulation avoids the singularity
issue that marks the Big Bang scenario, maintaining a
self-consistent behavior for modeling non-singular cosmic evolution.
Moreover, the parameter $h$ is pivotal in governing behavior of the
dynamical features around the bounce, thereby enabling the model
ability to describe a wide range of physical phenomena and remains
compatible with observational evidence.
\begin{figure}[ht]
\centering
\includegraphics[width=0.7\textwidth]{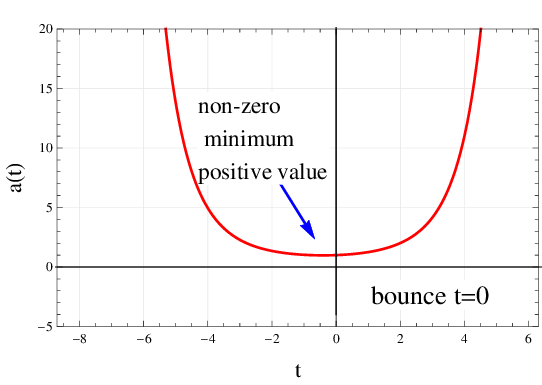}
\caption{Behaviour of the scale factor over cosmic time}
\end{figure}

Figure \textbf{1} presents the profile of the scale factor. The
scale factor remains finite and non-zero throughout the cosmic
evolution and attains its minimum value at the bounce point ($t=0$),
confirming the absence of an initial singularity. The curve exhibits
a smooth and symmetric transition from the contracting phase ($t<0$)
to the expanding phase ($t>0$). The parametric value of $h$ ensures
a gradual expansion rate around the bounce, leading to a regular and
continuous cosmological evolution without any singular behavior.
\begin{figure}[ht]
\centering
\includegraphics[width=1.0\textwidth]{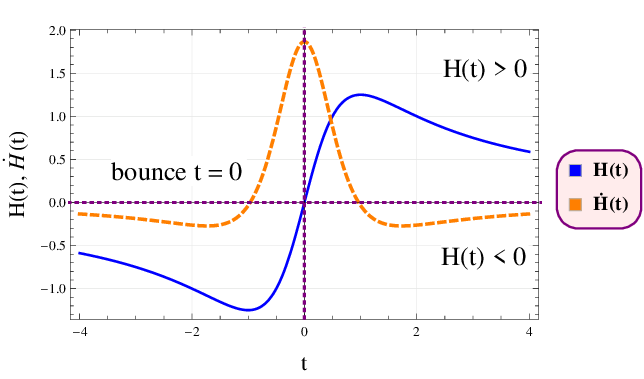}
\caption{Plot of the Hubble parameter and its time derivative.}
\end{figure}

In addition, the corresponding Hubble parameter fulfills the
required bounce conditions, allowing the universe to evolve
continuously from a contracting era ($t<0$) to an expanding era
($t>0$). Using Eq.(\ref{23}), the Hubble parameter and its time
derivative are obtained as
\begin{equation}\label{24}
H(t)=\frac{ht}{1+t^{2}}, \qquad
\dot{H}(t)=\frac{h(1-t^{2})}{(1+t^{2})^{2}}.
\end{equation}
At the bounce corresponding to $t=0$, the Hubble parameter and its
time derivative satisfy $H(0)=0$ and $\dot{H}(0)=h>0$, respectively,
thereby fulfilling the fundamental criteria for a non-singular
cosmic bounce. Figure \textbf{2} depicts the behavior of $H(t)$ and
$\dot{H}(t)$ in the neighborhood of the bounce. The Hubble parameter
evolves from negative values before $t=0$ to positive values
afterward, indicating a smooth transition from cosmic contraction to
expansion. Meanwhile, $\dot{H}(t)$ remains positive around the
bounce, further confirming the required dynamical condition. Hence,
the adopted scale factor provides a regular and continuous bouncing
evolution while avoiding a cosmological singularity.

Solving the continuity equation $\dot{\rho}+3H(\rho+p)=0$ for the
modified Chaplygin gas yields the energy density as a function of
the scale factor in the form
\begin{equation}\label{25}
\rho(a)=\bigg[\frac{A}{1+B}+\frac{C}{a^{3(1+\gamma)(1+B)}}\bigg]^{\frac{1}{1+\gamma}}.
\end{equation}
Here, $C$ is the integration constant. Upon inserting the chosen
scale factor ansatz, the corresponding energy density is obtained as
\begin{equation}\label{26}
\rho(t)=\bigg[\frac{A}{1+B}+\frac{C}{a_{0}^{3(1+\gamma)(1+B)}
(1+t^{2})^{\frac{3h}{2}(1+\gamma)(1+B)}}\bigg]^{\frac{1}{1+\gamma}}.
\end{equation}
Also, by inserting \eqref{26} in \eqref{15} and by rearranging, the
the pressure becomes
\begin{eqnarray}
p(t)&=&A\bigg[\frac{A}{1+B}+\frac{C}{a_{0}^{3(1+\gamma)(1+B)}
(1+t^{2})^{\frac{3h}{2}(1+\gamma)(1+B)}}\bigg]^{\frac{1}{1+\gamma}}\nonumber\\
&-&B\bigg[\frac{A}{1+B}+\frac{C}{a_{0}^{3(1+\gamma)(1+B)}
(1+t^{2})^{\frac{3h}{2}(1+\gamma)(1+B)}}\bigg]^{-\frac{\gamma}{1+\gamma}}.\label{27}
\end{eqnarray}
Using Eqs.\eqref{8}, \eqref{15}, \eqref{26} and \eqref{27} in
Eqs.\eqref{12}, \eqref{13}, we obtain
\begin{eqnarray}
\rho_{\rm eff}&=& \frac{ \left[ \dfrac{A}{1+B} + \dfrac{C}{
a_0^{3(1+\gamma)(1+B)}
\left(1+t^2\right)^{\frac{3h}{2}(1+\gamma)(1+B)} }
\right]^{\frac{1}{1+\gamma}} }{ 2\left[ 1+\psi n \left(
\dfrac{6h^2t^2}{(1+t^2)^2} \right)^{n-1} \right] }
\nonumber\\[3mm]
&+&\frac{\frac{1}{2}\bigg[\dfrac{6h^2t^2}{(1+t^2)^2}+\psi\bigg
(\dfrac{6h^2t^2}{(1+t^2)^2}\bigg)^{n}\bigg]}{ 2\left[ 1+\psi n
\left( \dfrac{6h^2t^2}{(1+t^2)^2} \right)^{n-1} \right] }.\label{28}
\end{eqnarray}
\begin{eqnarray}
p_{\rm eff} &=& -\Bigg[ B \left( \frac{A}{1+B} +\frac{C}
{a_0^{3(1+\gamma)(1+B)} (1+t^2)^{\frac{3h}{2}(1+\gamma)(1+B)}}
\right)^{\frac{1}{1+\gamma}}
\nonumber\\[2mm]
&\times& \frac{1}{2\left[ 1+\psi n \left( \dfrac{6h^2t^2}{(1+t^2)^2}
\right)^{n-1} \right]}+\frac{1}{2} \left[ \frac{6h^2t^2}{(1+t^2)^2}
+\psi \left(\frac{6h^2t^2}{(1+t^2)^2} \right)^n \right]
\nonumber\\[2mm]
&-&A\left(\frac{A}{1+B} +\frac{C} {a_0^{3(1+\gamma)(1+B)}
(1+t^2)^{\frac{3h}{2}(1+\gamma)(1+B)}}
\right)^{-\frac{\gamma}{1+\gamma}}\Bigg]
\nonumber\\[3mm]
&+&\frac{ 12\psi n(n-1) \left( \dfrac{h^2t^2}{(1+t^2)^2} \right)
\left( \dfrac{6h^2t^2}{(1+t^2)^2} \right)^{n-2} \left(
\dfrac{h(1-t^2)}{(1+t^2)^2} \right)} { 1+\psi n \left(
\dfrac{6h^2t^2}{(1+t^2)^2} \right)^{n-1} }.\label{29}
\end{eqnarray}
We evaluate the effective \textit{EoS}
$\omega_{eff}=\frac{p_{eff}}{\rho_{eff}}$ as
\begin{eqnarray}
\omega_{eff}(t) &=& \bigg[-BR + AR^{-\gamma} -\dfrac{1}{2}\left[
\dfrac{6h^{2}t^{2}}{(1+t^{2})^{2}}
+\psi\left(\dfrac{6h^{2}t^{2}}{(1+t^{2})^{2}}\right)^{n} \right]
\nonumber\\[3mm]
&+&\dfrac{24\psi n(n-1)h^{3}t^{2}(1-t^{2})}{(1+t^{2})^{4}}
\left(\dfrac{6h^{2}t^{2}}{(1+t^{2})^{2}}\right)^{n-2}\bigg]\Bigg/
\nonumber\\[3mm]
&&\bigg[-R +\dfrac{1}{2}\left[ \dfrac{6h^{2}t^{2}}{(1+t^{2})^{2}}
+\psi\left(\dfrac{6h^{2}t^{2}}{(1+t^{2})^{2}}\right)^{n}
\right]\bigg], \label{30}
\end{eqnarray}
where
\begin{equation}
R=\left( \frac{A}{1+B} +\frac{C} {a_0^{3(1+\gamma)(1+B)}
(1+t^2)^{\frac{3h}{2}(1+\gamma)(1+B)}}
\right)^{\frac{1}{1+\gamma}}.\label{31}
\end{equation}

Figure \textbf{3} shows the evolution of the effective \textit{EoS}
$\omega_{eff}$ for different positive values of the geometry
coupling parameter $\psi$. In all cases, $\omega_{ eff}$ exhibits a
similar evolution around the bounce. During the contracting phase
($t<0$), the \textit{EoS} parameter initially decreases from values
close to $\omega_{eff}=-1$ and reaches a minimum below the phantom
divide. It then increases as the universe approaches the bounce,
attaining positive values at $t=0$. A corresponding behavior is
observed on the other side of the bounce, reflecting the symmetric
nature of the adopted bouncing scenario.
\begin{figure}[ht]
\centering
\includegraphics[width=0.8\textwidth]{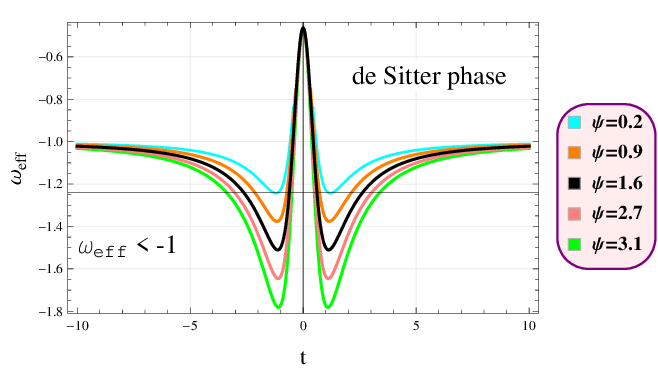}
\caption{Cosmic evolution of \textit{EoS} parameter}
\end{figure}

The evolution of $\omega_{eff}$ indicates a significant change in
the effective cosmic dynamics near the bounce. In particular, its
transition from the phantom regime demonstrates that the effective
fluid properties evolve rapidly in the bounce epoch, entering the de
Sitter regime. This behavior suggests that the cosmos approaches to
a de Sitter-like regime at late times. As is evident from Figure
\textbf{3} that the overall evolution of $\omega_{ eff}$ remains
qualitatively the same for distinct values of $\psi$, while the
extent of its deviation near the bounce varies with the selected
value of the coupling parameter. This shows that the polynomial
correction $\psi Q^{n}$ influences the effective cosmological
dynamics without removing the non-singular bouncing behavior. Hence,
the \textit{EoS} analysis further supports the robustness of the
bouncing solution within the considered $\mathcal{F}(Q)$ model.

\subsubsection{Evolution of Energy Conditions}

The \textit{ECs}, formulated from the energy-momentum tensor,
constitute essential theoretical constraints in gravitational
physics and cosmology. These laws dictate the allowable behavior of
the stress-energy tensor for matter sources by setting restrictions
on the allowed ranges of energy density and pressure due to the
gravitational attraction. Four standard forms are commonly employed:
\textit{NEC}, the strong energy condition (\textit{SEC}), the weak
energy condition (\textit{WEC}) and the dominant energy condition
(\textit{\textit{DEC}}), whose definitions are summarized in Table
\textbf{1} \cite{46}. These conditions provide a framework for
testing the internal consistency of cosmological scenarios and for
characterizing the matter content associated with specific
geometries. In this work, the \textit{ECs} are examined graphically
for the $\mathcal{F}(Q)$ model to assess their influence on cosmic
evolution. Notably, violation of the \textit{NEC} automatically
entails violation of the remaining \textit{ECs} \cite{47}. By
evaluating the \textit{ECs} for different choices of coupling
parameter $\psi$, the present analysis explores how the physical
properties of matter and geometry are inter-related within such
models.
\begin{table}\caption{\textbf{Classification of Energy Conditions}}
\begin{center}
\begin{tabular}{|c|c|}
\hline \textbf{Energy Conditions} & \textbf{Perfect Fluid}
\\
\hline \textit{NEC}   & $\rho_{\rm eff}+p_{\rm eff}\geq0$
\\
\hline\textit{SEC}   & $\rho_{\rm eff}+3p_{\rm eff}$, $\rho+p\geq0$
\\
\hline \textit{DEC}  &  $\rho_{\rm eff}\geq|p_{\rm eff}|$
\\
\hline  \textit{WEC}  & $\rho_{\rm eff}\geq0$, $\rho_{\rm
eff}+p_{\rm eff}\geq0$
\\
\hline
\end{tabular}
\end{center}
\end{table}
\begin{figure}\center
\epsfig{file=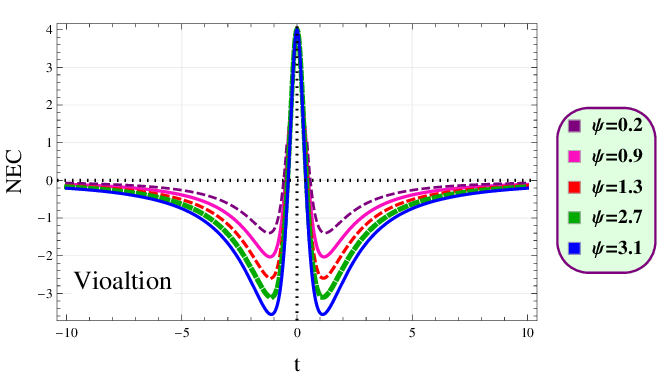,width=0.6\linewidth}
\epsfig{file=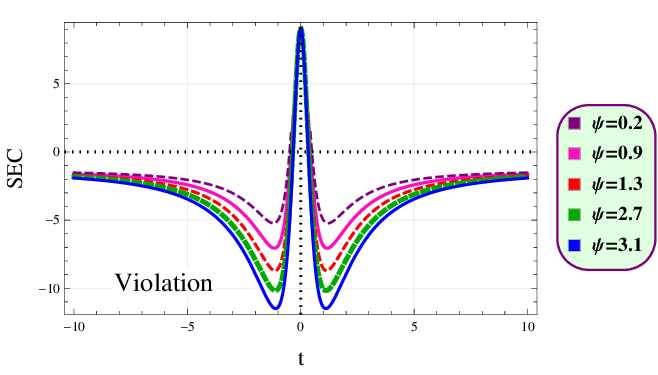,width=0.6\linewidth}
\epsfig{file=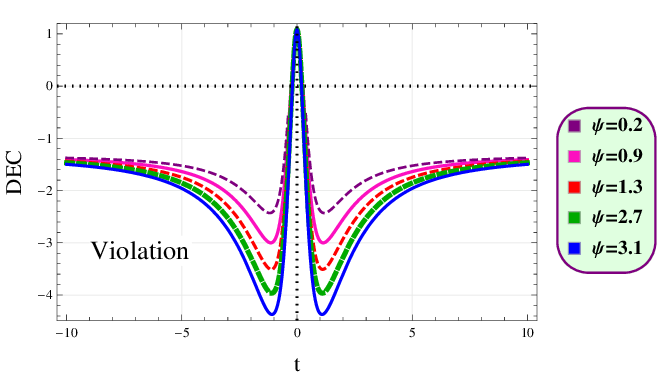,width=0.6\linewidth}
\epsfig{file=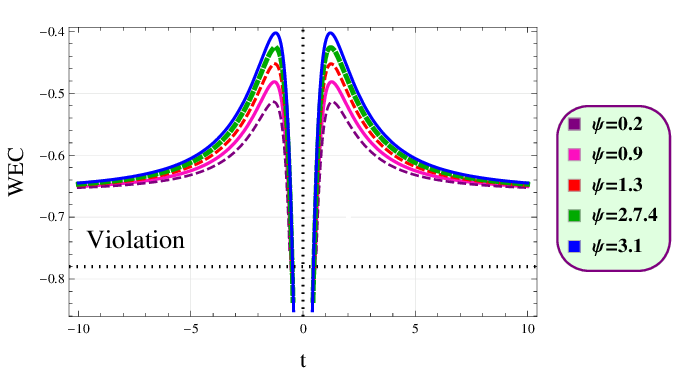,width=0.6\linewidth} \caption{Behavior of
\textit{ECs} corresponding to cosmic time for Model I.}
\end{figure}

The \textit{ECs} provide an important test of the physical behavior
of the bouncing solution. Figure~\textbf{4} shows the evolution of
the effective \textit{ECs} for Model I. In the vicinity of the
bounce, \textit{NEC} is violated, which is consistent with the
requirement $\dot{H}>0$ for a transition from contraction to
expansion. The remaining \textit{ECs} also exhibit departures from
their standard bounds around the bouncing epoch, reflecting the
strong influence of the modified gravitational sector. This behavior
indicates that the non-singular transition is supported by effective
geometric contributions arising from the $\mathcal{F}(Q)$
modification, rather than by introducing an exotic matter component.
The violation of the \textit{NEC} at high densities becomes a useful
regulator which prevents the singularity from occurring and ensures
a smooth bounce to the current expanding period.

\subsubsection{Stability and Causality Analysis}

For evaluating the viability of the bouncing solutions constructed
in $\mathcal{F}(Q)$ gravity, the speed of sound can be obtained
using the time derivatives of the effective energy density and
effective pressure as
\begin{equation}\label{33}
C_{s}^{2} = \frac{\dot{p}_{\rm eff}}{\dot{\rho}_{\rm eff}}.
\end{equation}
For a physically viable cosmological model, the squared sound speed
is generally required to satisfy
\begin{equation}\label{34}
0\leq C_{s}^{2}\leq1,
\end{equation}
where $C_{s}^{2}\geq0$ ensures the Laplacian instabilities on small
scales and $C_{s}^{2}\leq1$ guarantees that the propagation speed of
perturbations will not propagate faster than the speed of light.
Figure \textbf{5} shows the profile of the evolution of $C_s^2$
around the bounce for positive values of $\psi$. The behavior of the
trajectories can be used to examine the stability and causality of
the bouncing solutions. The trajectory remains within the physically
admissible interval $0\leq C_s^2\leq1$, indicating the absence of
Laplacian instabilities and superluminal propagation of
perturbations. A noticeable variation in $C_s^2$ occurs around the
bouncing epoch, reflecting the strong influence of the modified
gravitational sector in the high-density regime. At late times,
$C_s^2$ approaches a value close to the conformal limit,
$C_s^2=1/3\simeq0.333$, indicating a regular asymptotic behavior of
the effective cosmic fluid. Overall, the evolution of $C_s^2$ within
the stability and causality bounds supports the physical viability
of the proposed $\mathcal{F}(Q)$ bouncing model.
\begin{figure}[ht]
\centering
\includegraphics[width=1.0\textwidth]{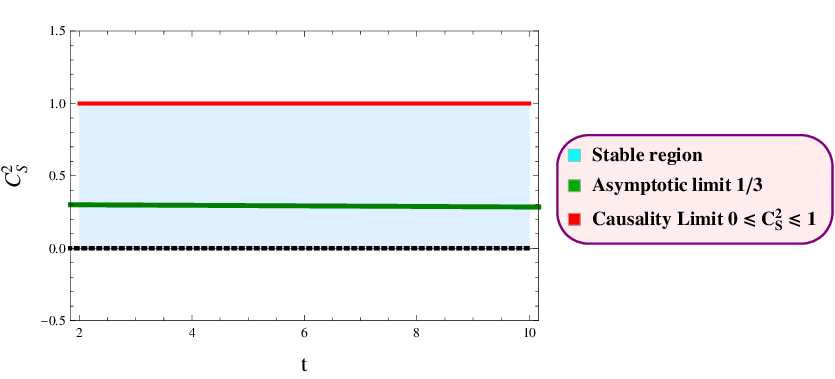}
\caption{Evolution of the squared sound speed $C_s^2$. }
\end{figure}

\subsection{Model II: Dynamical Evolution in the Autonomous System}

In order to examine whether the bounce comes about intrinsically due
to the $\mathcal{F}(Q)$ theory itself or due to the particular scale
factor chosen, we rewrite the modified Friedmann equations into a
two-dimensional autonomous dynamical system in the $(H,\rho)$ phase
space. The dynamical equations are derived using the Raychaudhuri
and continuity equations. Using the modified Chaplygin gas
\textit{EoS}, the autonomous system is expressed as
\begin{eqnarray}
\dot{H}&=&-\frac{1}{2} \bigg[\frac{1}{2\left(1+n\psi
Q^{\,n-1}\right)}\bigg[\rho+\bigg(A\rho-\frac{B}{\rho^{\gamma}}\bigg)\bigg]+2n(n-1)H\psi
Q^{\,n-2}\bigg],\label{35}\\
\dot{\rho}&=&-3H\bigg[\rho+\bigg(A\rho-\frac{B}{\rho^{\gamma}}\bigg)\bigg].\label{36}
\end{eqnarray}

The autonomous dynamical equations are solved numerically to obtain
the time evolution of $H(t)$ and $\rho(t)$ for fixed values of the
model parameters and suitable initial conditions. The scale factor
is then reconstructed numerically by integrating $\dot{a}(t)=H a(t)$
with an appropriate initial condition for $a(t)$. Since the
autonomous system is solved numerically, the reconstructed scale
factor is obtained in numerical form rather than as a closed
analytical expression. The numerical evolution of $a(t)$ is then
used to study the occurrence of the non-singular bounce regardless
of the initial prescribed value of the scale factor. The existence
of a minimum value for $a(t)$ along with the change in sign of the
Hubble parameter from $H(t)<0$ to $H(t)>0$ shows that there is a
smooth transition between the contracting era and the expanding era.
\begin{figure}[ht]
\centering
\includegraphics[width=0.7\textwidth]{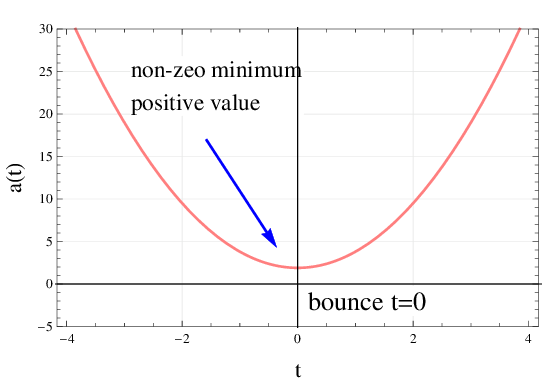}
\caption{Plot of the scale factor for Model II}
\end{figure}

Figure \textbf{6} shows the dynamics of the $a(t)$ for positive
value of the non-metricity coupling constant $\psi=1.6$. One can see
that the dynamics of the Model have a finite minimum value of the
scale factor at the point of bounce, then smoothly passes into the
stage of expansion. The stability of such a behavior suggests that
the non-singular transition does not depend on the specific form of
the ansatz for the scale factor. Furthermore, the symmetric
evolution of the scale factor on both sides of the bounce confirms
the robustness of the bouncing solution in the chosen
$\mathcal{F}(Q)$ setup for Model II.

\subsubsection{Analysis of \textit{EoS} Parameter and Squared Speed of Sound}

A further indicator of the dynamical evolution of the autonomous
system is the effective \textit{EoS} parameter, defined as
\begin{equation}
\omega_{eff}=\frac{p_{\rm eff}}{\rho_{\rm eff}}.
\end{equation}
Unlike the reconstruction approach, where the effective \textit{EoS}
is obtained from a prescribed scale factor, the present analysis
determines $\omega_{eff}$ directly from the numerical solutions of
the autonomous system. Starting from an initial contracting
configuration with $H_{0}<0$ and $\rho_{0}>0$, the coupled
differential equations for $H(t)$ and $\rho(t)$ are solved
numerically. The corresponding values of $p_{\rm eff}$ and
$\rho_{\rm eff}$ are then evaluated along the numerical trajectories
to determine the evolution of $\omega_{eff}$.
\begin{figure}[ht]
\centering
\includegraphics[width=0.6\textwidth]{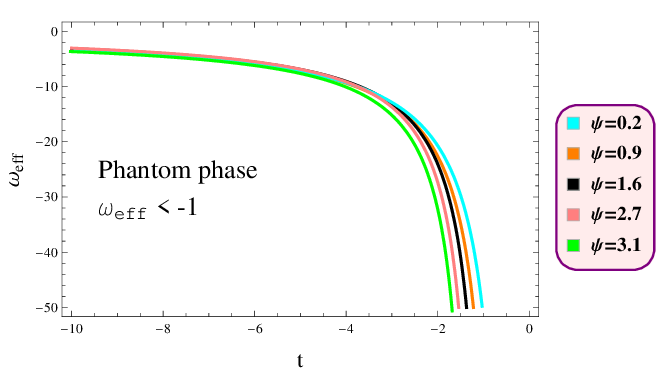}
\caption{Evolution of \textit{EoS} parameter for Model II}
\end{figure}

Figure \textbf{7} displays the behavior of $\omega_{eff}$ for the
positive $\psi$. In all cases, the effective \textit{EoS} begins
close to the cosmological-constant value, $\omega_{eff}\simeq-1$,
and decreases as cosmic time increases. The evolution towards values
below $\omega_{eff}=-1$ indicates the development of an effective
phantom-like regime during the subsequent evolution. This behavior
reflects the influence of the modified gravitational sector on the
effective pressure-to-density relation and demonstrates that the
effective cosmic fluid cannot be described by a conventional matter
component alone.

The qualitative evolution of $\omega_{eff}$ for Model II remains
consistent for all the considered positive values of $\psi$,
although its quantitative behavior is modified as the non-metricity
coupling is increased. This indicates that the correction term $\psi
Q^{n}$ influences the effective cosmological dynamics while
preserving the overall character of the solution. In particular, the
persistence of the same \textit{EoS} trend for distinct values of
$\psi$ demonstrates that the observed behavior is not restricted to
a single choice of the coupling parameter.

The decrease of $\omega_{eff}$ from values near $-1$ towards the
phantom regime is consistent with the strong modification of the
effective cosmic dynamics in the vicinity of the bounce. Since the
bounce requires a departure from the standard attractive-gravity
behavior, the evolution of the effective \textit{EoS} provides
complementary evidence for the role of the non-metricity
contribution in supporting the bouncing cosmology. Thus, the
numerical \textit{EoS} analysis provides a consistent description of
the changing dynamical behavior of the non-singular solution in the
considered $\mathcal{F}(Q)$ gravity framework.

In addition, we evaluate the squared speed of sound ($C_s^2
=\frac{\dot{p}_{\rm eff}}{\dot{\rho}_{\rm eff}}$) to further examine
the physical viability and dynamical stability of the Model II.
Figure \textbf{8} depicts the evolution of $C_s^2$ for the
considered positive value $\psi=1.6$. The stability and causality of
the Model are assessed through the conditions $0\leq C_s^2\leq1$,
where the lower bound ensures the absence of Laplacian
instabilities, while the upper bound guarantees subluminal
propagation of perturbations. As shown in Figure \textbf{8}, the
trajectories remain within the stable and causal interval throughout
the relevant cosmic evolution. This confirms that the bouncing
solutions remain free from small-scale Laplacian instabilities while
the propagation of perturbations remains causal.
\begin{figure}[ht]
\centering
\includegraphics[width=1.0\textwidth]{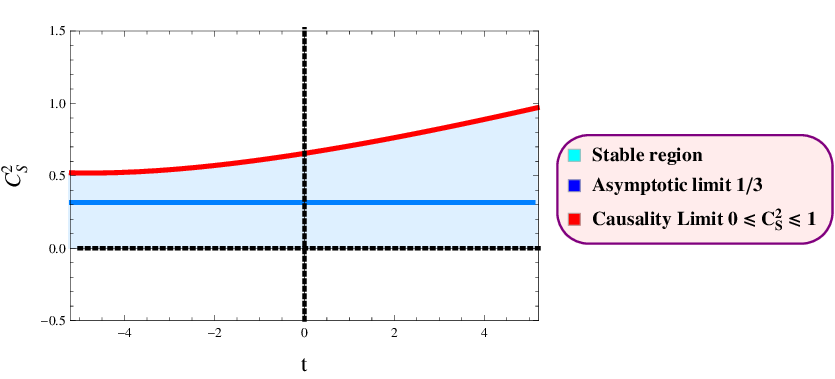}
\caption{Stability and causality scan for Model II.}
\end{figure}

During the evolution of the universe beyond the bounce, the quantity
$C_s^2$ tends towards $1/3$. The behavior of $C_s^2$ is compatible
with the relativistic sound speed for ultra-relativistic matter. The
convergence towards this value further indicates that the effective
fluid approaches a well-behaved regime at late times. These results
demonstrate that the inclusion of the $\psi Q^n$ term does not
compromise the stability or causality of the bouncing solutions and
provides further support for the physical viability of the
$\mathcal{F}(Q)$ cosmology.

\subsubsection{Phase Space Analysis}

Phase-space analysis provides an important dynamical test of the
non-singular bouncing solution. In contrast to the
scale-factor-based reconstruction, where the bounce is incorporated
through a prescribed ansatz, the autonomous-system formulation
allows the cosmological evolution to be examined directly from the
$\mathcal{F}(Q)$ field equations. The phase portrait in the
$(\rho,H)$ plane traces the trajectory from the contracting regime
($H<0$) through the bounce surface ($H=0$) to the expanding regime
($H>0$). A smooth crossing of $H=0$ without a divergence in the
energy density demonstrates that the bounce is dynamically supported
by the modified gravitational dynamics. Furthermore, the phase-space
trajectories provide insight into the robustness of the bouncing
solution and its subsequent cosmological evolution.
\begin{figure}[ht]
\centering
\includegraphics[width=1.0\textwidth]{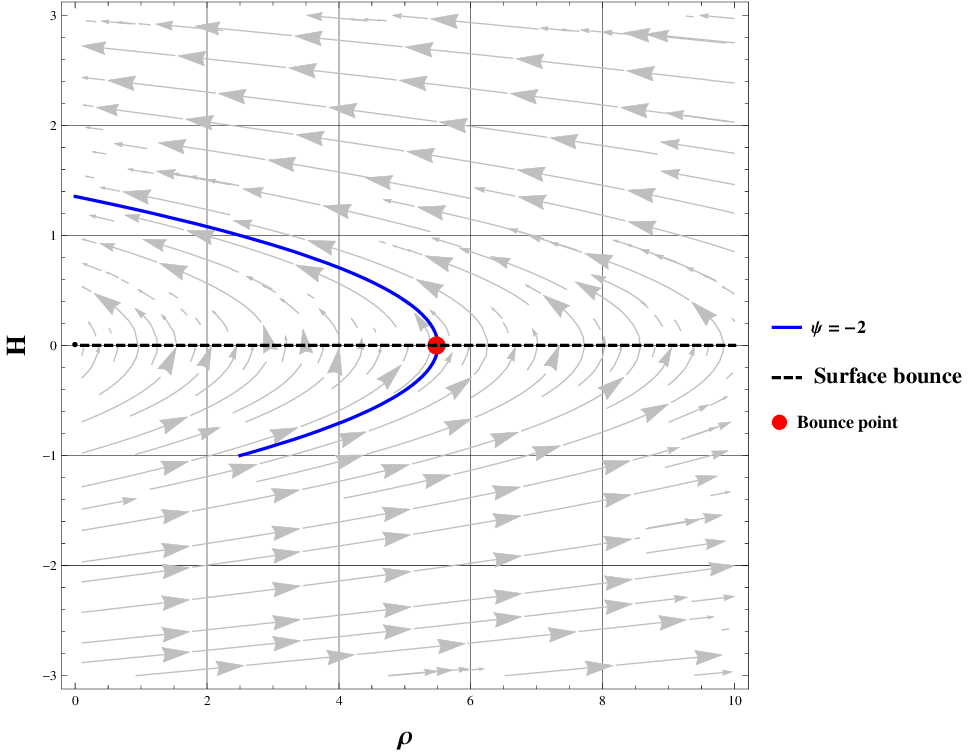}
\caption{Phase-space profile of the autonomous system in the
$(\rho,H)$ plane.}
\end{figure}

We numerically integrate the autonomous system by considering an
initial contracting region with $H_{0}<0$. Figure \textbf{9} depicts
the corresponding vector field and numerical trajectories in
$(\rho,H)$ phase plane. Stream lines represent the dynamics of the
flow in the universe, starting from the contraction regime $(H<0)$,
reaches $H=0$ and finally moves to the expansion phase $(H>0)$. The
trajectory passes smoothly through the $H=0$ boundary, indicating
the occurrence of a non-singular bounce rather than an evolution
towards a singular state. The phase-space behavior also provides
insight into the role of the non-metricity correction in the
bouncing dynamics. In Model II, the parameter $\psi$ controls the
contribution of the nonlinear non-metricity term to the modified
gravitational dynamics. For the considered positive values of
$\psi$, the numerical trajectory preserves the characteristic
contraction-to-expansion transition. Thus, the bouncing behavior
does not depend on a particular value of the coupling parameter but
persists across the range of positive $\psi$ investigated in this
work.

To further examine the robustness of the bouncing solution, we
perform a numerical scan of the $(\rho_0,\psi)$ parameter space,
where $\rho_0$ denotes the initial energy density and $\psi$
characterizes the strength of the non-linear non-metricity
correction. This analysis identifies the regions of parameter space
that lead to a successful transition from contraction to expansion.
The parametric map also illustrates the dependence of the bounce on
the initial cosmological state and the strength of the
$\mathcal{F}(Q)$ modification.
\begin{figure}[ht]
\centering
\includegraphics[width=1.0\textwidth]{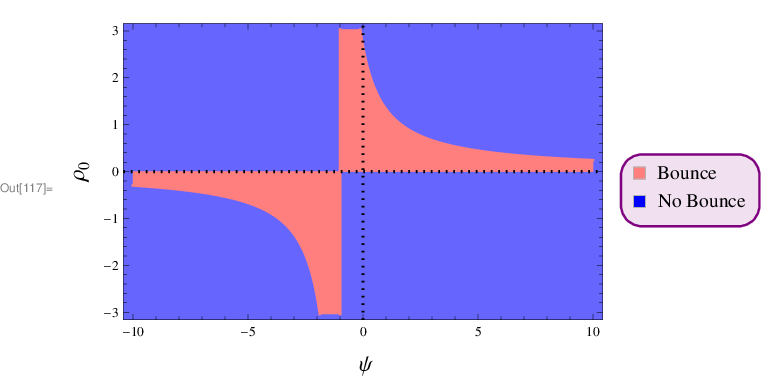}
\caption{Numerical profile of the parameter space
($\rho_0$,$\psi$).}
\end{figure}

Figure \textbf{10} illustrates the numerical parameter-space
analysis in the $(\psi,\rho_0)$ plane, where the blue region
represents bouncing regime and the orange region corresponds to the
no-bounce regime. The distribution of these regions shows that the
presence of a non-singular bounce depends on both the positive
non-metricity coupling parameter $\psi$ and the initial energy
density $\rho_0$. In particular, the presence of a distinct bouncing
region indicates that sufficiently suitable values of $\psi$ and
$\rho_0$ allow the universe to undergo a smooth transition from
contraction to expansion. The persistence of the bouncing region
over a range of parameter values further demonstrates that the
obtained non-singular behavior is not associated with a particular
parameter choice, supporting the robustness of the bouncing scenario
in the $\mathcal{F}(Q)$ framework.

\section{Comparative Analysis and Physical Cosmic Interpretation}

The consistency between the results obtained from Model I and Model
II provide complementary support for the bouncing scenario in the
present $\mathcal{F}(Q)$ cosmology. Model I, based on the prescribed
scale factor, describes the temporal evolution of the cosmological
quantities explicitly, whereas Model II employs an autonomous
dynamical system to investigate the bouncing behavior directly from
the modified field equations. The agreement between these two
approaches indicates that the obtained bounce is not solely
associated with the selected scale factor ansatz.

The main complementary features of the two approaches can be
summarized as follows.
\begin{itemize}
\item \textbf{Resolution of the Singularity:} Model I demonstrates a
continuous contraction-to-expansion change through the chosen scale
factor, which possesses a finite minimum value approximately
$a_{\min}\simeq1.2$ at the bounce. In contrast, the numerical
reconstruction obtained from the autonomous system in Model II gives
a comparatively larger minimum value, $a_{\min}\simeq1.9$ but the
same qualitative transition is recovered from the autonomous
equations for $H$ and $\rho$, providing an independent dynamical
indication that the non-singular bounce is supported by the field
equations rather than being exclusively imposed through the scale
factor. This agreement between the two approaches indicates that the
singularity avoidance is consistently supported within the
$\mathcal{F}(Q)$ framework.

\item \textbf{Role of the Non-metricity Coupling:} In both approaches,
the non-linear term $\psi Q^n$ modifies the cosmic dynamics in the
vicinity of the bounce. The persistence of the bouncing behavior for
the considered positive values of $\psi$ demonstrates that the
transition is not restricted to a particular value of the coupling
parameter.

\item \textbf{Effective Equation of State:}
The two models exhibit distinct but physically meaningful evolutions
of the effective \textit{EoS} parameter. In Model I, $\omega_{eff}$
initially decreases from $-1$ into the phantom regime and
subsequently increases towards the de Sitter value,
$\omega_{eff}\rightarrow -1$, at late times. In contrast, Model II
crosses the phantom divide $\omega_{eff}=-1$ and subsequently
remains within the phantom regime during the considered evolution.
This difference reflects the distinct dynamical characteristics of
the two approaches but the similar qualitative behavior in both
cases indicates that the effective cosmic dynamics are consistently
described within the $\mathcal{F}(Q)$ context.

\item \textbf{Stability and Causality:} The squared speed of sound
satisfies the physically required conditions $0\leq C_s^2\leq1$ for
the considered solutions. This indicates the absence of small-scale
Laplacian instabilities and ensures causal propagation of
perturbations. The convergence of the sound-speed behavior towards
the relativistic value $C_s^2=1/3$ at late times further supports
the physical viability of the solutions.

\item \textbf{Dynamical Origin of the Bounce:} Model I establishes
the bouncing behavior through a prescribed scale-factor evolution.
The phase-space trajectories and $(\rho_0,\psi)$ parameter-space
analysis in Model II further show that the bounce persists over
initial conditions and positive coupling strengths, supporting its
dynamical robustness.
\end{itemize}

The comparison between the two approaches demonstrates that the
bouncing behavior persists when the cosmological dynamics are
examined from both a prescribed scale-factor description and an
independent autonomous-system formulation. Table \textbf{2} presents
comprehensive comparison of both models.
\begin{table}[h]
\centering \caption{Comparison of Model I and Model II in
$\mathcal{F}(Q)$ Gravity.} \label{2}
\begin{tabular}{|l|c|c|}
\hline
\textbf{Feature} & \textbf{Model I} & \textbf{Model II} \\
\hline
Primary drivers & Scale Factor Geometry & Phase Space Analysis\\
Minimum scale factor & $a_{\min}\simeq1.2$
& $a_{\min}\simeq1.9$ \\
Bounce & Non-singular & Non-singular \\
Bounce condition & $H=0,\ \dot{H}>0$ & $H=0,\ \dot{H}>0$ \\
Latetime \textit{EoS}  & de Sitter phase & phantom phase \\
Stability & $0\leq C_s^2\leq1$ & $0\leq C_s^2\leq1$ \\
Bouncing Source & Geometric $\psi Q^{n}$ & Geometric $\psi Q^{n}$ \\
\hline
\end{tabular}
\end{table}

\subsection{Physical Aspects of the $\mathcal{F}(Q)$ Bounce}

The results produced from this study indicate that the non-linear
non-metricity sector can significantly modify the gravitational
dynamics in the high-density regime and provide a mechanism for
avoiding the cosmological singularity. In standard \textit{GR}, the
Raychaudhuri equation describes the focusing of nearby geodesics
under the influence of attractive gravity and ordinary matter. In
the present  $\mathcal{F}(Q)$ framework, however, the non-linear
contribution $\psi Q^n$ introduces additional geometric terms into
the modified gravitational equations. These corrections become
increasingly relevant as the magnitude of the non-metricity scalar
grows.

Near the bounce, the gravitational affects can counteract the usual
attractive behavior and permit the Hubble variable to shift from a
negative value to zero at the bounce spot and subsequently to
positive values during the expanding phase. Thus, non-linear
non-metricity contribution provides an effective geometric mechanism
capable of supporting the transition
\begin{equation}
H<0 \quad \longrightarrow \quad H=0 \quad \longrightarrow \quad H>0,
\end{equation}
while the scale factor remains finite and non-zero at the bounce.

An important aspect of the present scenario is that the required
modification of the gravitational dynamics does not necessarily
require the introduction of an explicitly exotic matter component.
The Chaplygin-type matter sector can retain its standard behavior,
while the effective departure from the usual attractive
gravitational dynamics arises from the non-linear geometric
contribution in  $\mathcal{F}(Q)$. In this sense, the singularity
avoidance is associated with the modified gravitational sector
rather than being imposed solely through an exotic matter source.

The persistence of the bouncing behavior for the values of the
coupling parameter $\psi$ further indicates that the geometric
mechanism is not restricted to a particular fine-tuned parameter
choice. Therefore, the non-linear $\mathcal{F}(Q)$ framework
provides a promising geometric setting in which the high-density
evolution of the universe can be modified sufficiently to replace
the conventional singular beginning with a smooth transition from
contraction to expansion.

\section{Summary}

In this work, we have investigated the possibility of resolving the
cosmological initial singularity through bouncing scenario in the
context of $\mathcal{F}(Q)$ theory. We have considered the
gravitational Lagrangian in the presence of a Chaplygin-type matter
source. The non-linear dependence on the non-metricity scalar
introduces additional geometric contributions to the gravitational
dynamics, which can become important in the high-density regime. To
examine the origin, evolution, and physical viability of the bounce,
we have employed two complementary approaches: Model I employs a
kinematic reconstruction based on a suitable scale-factor ansatz
whereas Model II is formulated within the context of an autonomous
dynamical-system. The main results of this study are presented
below.
\begin{itemize}

\item \textbf{Scale Factor Evolution:}
The scale factor for both cases is always finite and non-vanishing
at the bouncing point and smoothly changes from contraction phase to
the expanding one, thus confirming the robustness of the bouncing
scenario (Figures \textbf{1} (Model I) and \textbf{6} (Model II)).

\item \textbf{Hubble Parameter and Time Derivative:}
The Hubble parameter transitions smoothly from $H<0$ to $H>0$ across
$H=0$ at the point of bounce. Moreover, the value of $\dot{H}$
remains positive during the transition, which satisfies the critical
condition of non-singularity of a cosmological bounce (Figure
\textbf{2} (Model I)).

\item \textbf{Effective \textit{EoS}:}
The effective \textit{EoS} exhibits distinct evolutionary behavior
in both cases. In Model I, it enters the phantom regime and
subsequently approaches the de Sitter limit, $w_{\rm eff}\rightarrow
-1$ (Figure \textbf{3}), whereas in Model II, it remains in the
phantom regime after crossing the phantom divide, $w_{\rm eff}=-1$
(Figure \textbf{7}). This implies that $\mathcal{F}(Q)$ gravity
provides a framework to describe the history of the universe, from
avoiding the primordial singularity to the current era dominated by
dark energy.

\item \textbf{Energy Conditions:}
The effective \textit{NEC} is violated in the vicinity of the
bounce, providing the necessary condition for the transition from
contraction to expansion. The depletion of \textit{NEC} consequently
leads to the violation of all other \textit{ECs}, reflecting the
effective repulsive gravitational behavior required to facilitate
the non-singular bounce (Figure \textbf{4}).

\item \textbf{Sound-speed Stability:}
The squared sound speed remains within the physically acceptable
stability and causality range, $0\leq C_s^2\leq1$, throughout the
considered evolution (Figures \textbf{5} (Model I) and \textbf{8}
(Model II)). This behavior indicates the absence of Laplacian
instabilities and supports the dynamical viability of the bouncing
solutions.

\item \textbf{Phase Portrait Analysis:}
The phase-space trajectories are obtained from the autonomous
dynamical system. We have found that the trajectories smoothly cross
the $H=0$ boundary, connecting the contracting and expanding
branches, which confirm that the bounce is also reproduced
dynamically and is not solely a consequence of the chosen
scale-factor ansatz (Figure \textbf{9}).

\item \textbf{Parameter Space Analysis:}
The numerical scan of parameter space reveals a distinct region in
the $(\rho_0, \psi)$ plane where non-singular bouncing solutions are
obtained for Model-II. The existence of an extended bounce and
no-bounce regimes demonstrate that the non-singular transition is
not confined to a specific choice of the initial density or coupling
parameter, but persists across a finite range of the considered
parameter space. (Figure \textbf{10}).
\end{itemize}

Overall, our results indicate that the non-linear $\mathcal{F}(Q)$
model $\mathcal{F}(Q)=Q+\psi Q^n$, when combined with a
Chaplygin-type matter source, can provide a viable framework for
describing a non-singular cosmological bounce. The non-linear
non-metricity contribution modifies the high-density gravitational
dynamics in a manner that permits the transition from contraction to
expansion while maintaining a finite scale factor. The agreement
between the kinematic reconstruction and autonomous dynamical
analysis, together with the acceptable \textit{ECs}, effective
\textit{EoS} and sound-speed behavior, supports the robustness of
the proposed scenario. These results demonstrate that non-linear
symmetric teleparallel gravity offers a promising geometric
framework for exploring singularity avoidance and alternative early
universe dynamics beyond the standard \textit{GR} description.

Our results share several qualitative features with the previously
studied Myrzakulov-type $\mathcal{F}(\mathcal{R},\mathbb{T})$
bouncing model with Chaplygin gas \cite{39a}, particularly the
realization of a non-singular contraction-to-expansion transition
and the violation of the effective \textit{NEC} near the bounce. In
$\mathcal{F}(\mathcal{R},\mathbb{T})$ framework, the bounce is
primarily associated with the quadratic trace coupling $\beta
\mathbb{T}^{2}$, where negative $\beta$ provides the required
effective repulsive contribution, whereas in the present
$\mathcal{F}(Q)$ context, the bounce is driven by non-linear
non-metricity corrections with positive $\psi$. Moreover, our
autonomous analysis independently confirms the bouncing behavior
through phase-space and parameter-space evolution, providing a
complementary dynamical perspective on the non-singular cosmic
evolution.

It is useful to compare the present $\mathcal{F}(Q)$ bouncing
scenario with other well-known approaches to non-singular cosmology.
In Loop Quantum Cosmology, the classical cosmological dynamics are
modified by quantum-geometrical effects that become important near
the Planck-density regime, leading to a reversal from contraction to
expansion. In contrast, the present study realizes a similar
non-singular transition within a classical modified-gravity
framework through non-linear corrections in the non-metricity
sector. Thus, the bounce in our case does not require the explicit
incorporation of quantum gravitational corrections
\cite{51}-\cite{53}.

The present analysis opens several directions for future
investigations. An important extension would be to study the
evolution of cosmological perturbations \cite{54,55} and confront
the predictions of the model with observational data, particularly
Cosmic Microwave Background  observations \cite{56}. Such an
analysis would provide additional constraints on model parameters
and allow a more detailed comparison of the proposed bouncing
scenario with the standard inflationary picture. In particular, the
Relativistic Generalized Uncertainty Principle (RGUP) provides a
promising framework for examining possible quantum corrections to
the classical cosmological dynamics. Following the approaches
developed in related studies \cite{58}-\cite{63}, future work can
incorporate RGUP corrections into the present $\mathcal{F}(Q)$
framework to examine their influence on the
bouncing solution.\\\\
\textbf{Data Availability Statement:} No data was analyzed/produced
in this paper.

\end{document}